\documentclass{pasa}

\usepackage{aas_macros}
\usepackage{hyperref} 
\usepackage{graphicx}
\usepackage{url}
\usepackage{amsmath}
\usepackage{color}
 \usepackage{mathrsfs} 
 \usepackage{lscape,epsfig}
 \usepackage{amsmath}
 \usepackage{bm}
 \usepackage{txfonts}

  \def\beq{\begin{equation}}
  \def\eeq{\end{equation}}
  \def\beqa{\begin{eqnarray}}
  \def\eeqa{\end{eqnarray}}
  \def\ban{\begin{eqnarray*}}
  \def\ean{\end{eqnarray*}}
  \def\bi{\begin{itemize}}
  \def\ei{\end{itemize}}

\begin{document}

\title[Neutron star properties from optimized chiral nuclear interactions]
{Neutron star properties from optimized chiral nuclear interactions}
\author[Domenico Logoteta and Ignazio Bombaci]{ Domenico Logoteta$^1$, Ignazio Bombaci$^2$ 
\affil{$^1$INFN, Sezione di Pisa, Largo Bruno Pontecorvo 3, I-56127 Pisa, Italy}%
\affil{$^2$Dipartimento di Fisica, Universit\'a di Pisa, Largo Bruno Pontecorvo 3, I-56127 Pisa, Italy} }

\jid{PASA}
\doi{10.1017/pas.\the\year.xxx}
\jyear{\the\year}

\hypersetup{colorlinks,citecolor=blue,linkcolor=blue,urlcolor=blue}

\begin{frontmatter}
\maketitle

\begin{abstract}

We adopt two- and three-body nuclear forces derived at the next-to-next-to-leading-order (N2LO) in the framework 
of effective chiral perturbation theory (ChPT) to calculate the equation of state (EOS) of $\beta$-stable neutron star matter using the Brueckner--Hartree--Fock many-body approach. 
We use the recent optimized chiral two-body nuclear interaction at N2LO derived by \cite{ekstrom1} and 
two different parametrizations of the three-body N2LO interaction: 
the first one is fixed to reproduce the saturation point of symmetric nuclear matter while 
the second one is fixed to reproduce the binding energies of light atomic nuclei.   
We show that in the second case the properties of nuclear matter are not well determined 
whereas in the first case various empirical nuclear matter properties around the saturation density are 
well reproduced. 
We also calculate the nuclear symmetry energy $E_{sym}$ as a function of the nucleonic density  
and compare our results with the empirical constraints obtained using the excitation energies 
of isobaric analog states in nuclei and the experimental data on the neutron skin thickness of heavy nuclei.   
We next calculate various neutron star properties and in particular the mass-radius and mass-central density relations. We find that the adopted interactions based on a fully microscopic framework, are able to provide 
an EOS which is consistent with the present data of measured neutron star masses and in particular with the mass 
$M=2.01\pm0.04 M_\odot$ of the neutron star in PSR J0348+0432.  
We finally consider the possible presence of hyperons in the stellar core and we find a softening of the EOS 
and a substantial reduction of the stellar maximum mass in agreement with similar calculations present in the literature. 
\end{abstract} 

\end{frontmatter}

\section{Introduction}

The physics of neutron stars represents a great challenge to test our understanding of matter under extreme  conditions. The huge variation of the density from the star surface ($\rho \sim 10\, {\rm g/cm^3}$) to its 
center ($\rho \sim 10^{15}\, {\rm g/cm^3}$) requires the modeling of systems in very different physical conditions like heavy neutron rich nuclei arranged to form a lattice structure as in the outer crust of the star, 
or a system of strong interacting hadrons (nucleons, and possibly hyperons or a phase with deconfined quarks) 
to form a quantum fluid as in the stellar core \citep{prak97}.    
The description of such a variety of nuclear systems needs for a considerable theoretical effort 
and a knowledge as much as possible accurate of the interactions between the constituents present inside  
the star.  
The bulk properties of neutron stars (e.g. mass, radius, mass-shed frequency) chiefly depend on 
the equation of state (EOS) describing the macroscopic properties of stellar matter.  
The EOS of dense matter is also a basic ingredient for modeling various astrophysical phenomena related 
to neutron stars, as core-collapse supernovae (SNe) \citep{oertel2017} 
and binary neutron star (BNS) mergers  
\citep{BJ2012,Bernuzzi2015,Sekiguchi2016,RT2016}.  
We note however that in order to perform realistic numerical simulations for the latter two cases the inclusion of thermal contributions 
is very important. 
The very recent detection of gravitational waves from a binary neutron star merger (GW170817) by the LIGO-Virgo collaboration 
\citep{abbott5}, has strongly increased the interest to these astrophysical 
phenomena and more in general to to dense matter physics.  
   
In the present work we model the core of neutron stars as a uniform charge neutral fluid made of neutrons, protons, electrons and muons 
in equilibrium with respect to the weak interaction. Such system is well known in literature as $\beta$-stable nuclear matter. 
In addition we also consider the possible formation of hyperons in the inner core of neutron stars.   
Accordingly we calculate various neutron star properties making use of an EOS for the stellar core   
obtained within a microscopic non-relativistic approach based the Brueckner--Bethe--Goldstone (BBG) 
many-body theory and adopting the Brueckner--Hartree--Fock (BHF) approximation \citep{bbg1,bbg2}. 
In such a microscopic approach the only inputs required are the bare two- and three-body nuclear interactions  
derived in vacuum using nucleon-nucleon (NN) scattering data and informations (binding energies and scattering observables) on light (atomic mass number $A = 3$, 4) nuclei.   

It is well known that three-nucleon forces (TNFs) play a very important role in nuclear physics.   
For example, TNFs are required to reproduce the experimental binding energy of few-nucleon ($A = 3,\ 4$) 
systems \citep{kalantar12}. TNFs are also essential to reproduce the empirical saturation point 
($n_{0} = 0.16 \pm  0.01~{\rm fm}^{-3}$, $E/A|_{n_0} = -16.0 \pm 1.0~{\rm MeV}$) of symmetric nuclear matter (SNM) 
and to give an adequately stiff EOS which is consistent with present measured neutron star masses 
and in particular with the mass $M = 2.01\pm0.04 M_\odot$ (\cite{anto13}) of the neutron star in PSR J0348+0432.  

A modern and very powerful approach \citep{weinberg} to derive two- as well as many-body nuclear interactions is the one provided 
by chiral effective field theory (see\ \citep{epel09} and\ \citep{machl11} for a detailed review).   
In this method two-, three- as well as many-body nuclear interactions can be calculated order by order according to 
a well defined procedure based on a low-energy effective quantum chromodynamics (QCD) Lagrangian. This Lagrangian is built  
in such a way to keep the main symmetries of QCD and in particular the approximate chiral symmetry.   
The starting point of this chiral perturbation theory (ChPT) is the definition of a power counting in the ratio $Q/\Lambda_\chi$, 
where $Q$ denotes a low-energy scale wich can be identified with the momentum of the external  
nucleons or with the pion mass $m_\pi$. $\Lambda_\chi \sim 1~{\rm GeV}$ is the so called 
 chiral symmetry breaking scale which sets up the energy range of validity of the theory.  
In this effective field theory, the details of the QCD dynamics are enclosed in the so called low-energy 
constants (LECs), which are parameters fitted using experimental data such as scattering data and binding energies of light nuclei. 
This well defined scheme is very advantageous in the case of nucleonic systems where it has been shown that three-nucleon forces (TNFs) 
play a very important role (\cite{kalantar12}).   

In this work, we present some microscopic calculations of the EOS of $\beta$-stable neutron star matter 
using the chiral potentials derived by\ \cite{ekstrom1} at the next-to-next-to-leading-order (N2LO) of ChPT.     
Interactions derived in ChPT have been calculated even at higher order like N3LO and 
N4LO \citep{entem_n4lo,epelb_n4lo}. 
One of the problems to perform nuclear structure and nuclear matter calculations at a fixed order 
higher than N2LO, is that the number of many-body contributions proliferate very quickly increasing the order 
of the expansion.  
Therefore it turns out prohibitive to take into account all the contributions arising at a given arbitrary order 
of ChPT.  
Conversely at the order N2LO it has been shown by\ \cite{ekstrom1} that is possible to derive a NN potential 
with a $\chi^2/datum\sim\ 1$, as well as to take into account leading order TNFs. 
Previous versions of NN potentials at N2LO based on traditional fit techniques of the experimental data, 
provided a $\chi^2/datum \sim\ 10$ and therefore they were not enough accurate to be used in practical calculations.   
Alternatively\ \cite{ekstrom1} used a new optimization technique based on the algorithm   
POUNDerS (Practical Optimization Using No Derivatives for sum of Squares)\ \citep{pounders} which drastically improved 
the quality of the data fit. Thus at N2LO all the contributions emerging from ChPT can be consistently 
included in a many-body calculation.

\section{Chiral nuclear interactions}

As we have already discussed previously, in the present work we employ two different interactions derived in ChPT both 
for two and the three-body sectors. 
We adopt indeed a NN potential calculated at N2LO supplemented by a three-nucleon force calculated 
at the same order. 
More specifically as a two-body nuclear interaction, we have used the optimized chiral potentials proposed by\ \cite{ekstrom1}. 
We have already pointed out that all the possible operators contributing to the NN potential as well as   
leading order TNFs arise at N2LO of ChPT.   
Thus it is possible to understand several properties of nuclear structure at this order of the perturbative expansion.  
The optimized parameters of the NN potential fitted at N2LO are the   
constants $c_1$, $c_3$ and $c_4$ coming from the pion-nucleon ($\pi N$) Lagrangian, plus $11$ partial-waves from contact terms.  

The chiral NN interaction by\ \cite{ekstrom1} has been optimized to the proton-proton and the proton-neutron scattering data for laboratory scattering energies below $125\ {\rm MeV}$, and to deuteron observables. 
The N2LO TNF has been then fixed requiring to reproduce the $^{3}$H half-life and the binding energies  
of $^3$H and $^3$He nuclei.    
The total (i.e. two-body plus three-body) interaction has been then used to predict the Gamow-Teller transition 
matrix-elements in $^{14}$C and $^{22,24}$O nuclei using consistent two-body currents.   
In their paper\ \cite{ekstrom1} provided three different versions of this interaction according to three different 
values of the cutoff $\Lambda = 450, \, 500, \, 550$ ${\rm MeV}$ used to regularize the short range part 
of the potentials. 
The $\chi^2$/datum of the NN interaction varied from $1.33$ to $1.18$ passing from $\Lambda = 450$ to 
$\Lambda = 550$ ${\rm MeV}$.  
In the present work we have adopted the model with $\Lambda = 550$ ${\rm MeV}$ hereafter referred to as the 
N2LO$_{opt}$ NN potential.   
We have checked however that similar results could be obtained also using the other models 
reported in\ \cite{ekstrom1}. 

Concerning the form of the TNF, we have used the non-local N2LO version given by\ \cite{N2LO}.    
The non locality of the N2LO TNF depends only on the particular form of the cutoff used to regularize short range part 
the potential. It reads: 
\begin{equation}
V_{3N}^{(2\pi)} = \sum_{i\neq j\neq k} \frac{g_A^2}{8f_\pi^4} 
\frac{\bm{\sigma}_i \cdot \bm{q}_i \, \bm{\sigma}_j \cdot
\bm{q}_j}{(\bm{q_i}^2 + m_\pi^2)(\bm{q_j}^2+m_\pi^2)}
f_{ijk}^{\alpha \beta}\tau_i^\alpha \tau_j^\beta,
\label{nnn1}
\end{equation}
\begin{equation}
V_{3N}^{(1\pi)} = -\sum_{i\neq j\neq k}
\frac{g_A c_D}{8f_\pi^4 \Lambda_\chi} \frac{\bm{\sigma}_j \cdot \bm{q}_j}{\bm{q_j}^2+m_\pi^2}
\bm{\sigma}_i \cdot
\bm{q}_j \, {\bm \tau}_i \cdot {\bm \tau}_j ,
\label{nnn2}
\end{equation}
\begin{equation}
V_{3N}^{(\rm ct)} = \sum_{i\neq j\neq k} \frac{c_E}{2f_\pi^4 \Lambda_\chi}
{\bm \tau}_i \cdot {\bm \tau}_j,
\label{nnn3}
\end{equation}
where $\bm{q}_i=\bm{p_i}^\prime -\bm{p}_i$ is the difference between the 
final and initial momentum of nucleon $i$ and 
\begin{equation}
f_{ijk}^{\alpha \beta} = \delta^{\alpha \beta}\left (-4c_1m_\pi^2
 + 2c_3 \bm{q}_i \cdot \bm{q}_j \right ) + 
c_4 \epsilon^{\alpha \beta \gamma} \tau_k^\gamma \bm{\sigma}_k
\cdot \left ( \bm{q}_i \times \bm{q}_j \right ).
\label{nnn4}
\end{equation}
In equations (\ref{nnn1})--(\ref{nnn4}) $\bm\sigma_i$ and $\bm\tau_i$ are the Pauli matrices which act on the spin and 
isospin spaces while $g_A=1.29$ is the axial-vector coupling and $f_\pi = 92.4$ ${\rm MeV}$ the pion decay constant. 
The labels $i$, $j$, $k$ run over the values $1$, $2$, $3$, which take into account all the six possible 
permutations in each sum.   
In eq.\ \ref{nnn4} $c_1$, $c_3$, $c_4$, $c_D$ and $c_E$ denote the so called low energy constants.   
We note that $c_1$, $c_3$ and $c_4$ are already fixed at two-body level by the $\pi N$ 
Lagrangian, therefore they do not represent free parameters.     
In Tab.\ \ref{tab1} we report the values of $c_i$ that we have adopted in the present work. 
The last two parameters $c_D$ and $c_E$ are not fixed by the data from two-body scattering and have to be set up 
using some specific observable in finite nuclei or in infinite nuclear matter.  
In the present work we have explored both the possibilities.   
In the following of this paper the TNF fitted by\ \cite{ekstrom1} to reproduce the properties of light nuclei 
will be denoted as the N2LO TNF, whereas the parametrization fitted to provide a good saturation point of SNM 
will be denoted as the N2LO1 TNF.   

Finally, we have multiplied the whole interaction by a non local cut off of the form: 
\begin{equation}
F_\Lambda({\bm p}, {\bm q})={\rm exp}\left[-\left(\frac{4 {\bm p}^2+3 {\bm q}^2}{4 \Lambda^2}\right)^n\right] \;. 
\label{cut}
\end{equation}
This allows to regularize the short part of the interaction which is not correctly described by ChPT and it is sensible 
to the internal structure of nucleons. 
In Eq.\ \ref{cut}: ${\bm p}=({\bm p_1}-{\bm p_2)}/2$ and ${\bm q}=2/3[{\bm p_3}-({\bm p}_1-{\bm p}_2)]$.   
Finally, following\ \cite{ekstrom1}, in the present work we have set $\Lambda = 550\ {\rm MeV}$ and $n=2$.  

 \begin{table*} 
\begin{tabular}{l|cccccc}
\hline
      TNF model            & $c_D$ &   $c_E$  & $c_1$    & $c_3$   & $c_4$   \\               
\hline
  N2LO  & 0.1488  &   -0.747   & -0.906 & -3.897 & 3.906  \\ 
  N2LO1 &-0.5000  &    0.900   & -0.906 & -3.897 & 3.906  \\ 
\hline
\hline
 \end{tabular}
\caption{Values of the low energy constants (LECs) of the two TNF parametrizations used in the present work. 
For the two parametrizations we have set a cut off of $550$ ${\rm MeV}$. 
$c_D$ and $c_E$ are dimensionless whereas $c_1$, $c_3$ and $c_4$ are expressed in ${\rm GeV^{-1}}$. } 
\label{tab1}
\end{table*} 

\section{The BHF approach with three-body forces} 
\label{sec:2}

The Brueckner--Bethe--Goldstone (BBG) many-body theory \citep{bbg1,bbg2} allows to calculate the ground state 
of nuclear matter in terms of the so-called hole-line expansion. The different diagrams which contribute to the 
 energy of the system, are grouped according to the number of independent hole-lines, where  
the hole-lines represent empty single particle states in the Fermi sea. 
The lowest order the BBG theory is the so called Brueckner--Hartree--Fock (BHF) approximation. In the present work we have 
performed all the calculations in such framework.  
The starting point of the BHF approach is the calculation of the so called $G$-matrices which describe the interaction 
between two nucleons taking into account the presence of all the surrounding nucleons of the medium; these nucleons restrict 
the possible final states of the nucleon-nucleon scattering. 
 
For asymmetric nuclear matter with total nuclear density $\rho = \rho_n + \rho_p$ and isospin asymmetry 
$\beta = (\rho_n - \rho_p)/\rho$ , (being $\rho_n$ and $\rho_p$ 
the neutron and proton densities) one has to consider three different $G$-matrices for the $nn$-, $np$- and $pp$-channels. 
These $G$-matrices are obtained solving the well known Bethe--Goldstone equation:   
%
\begin{equation}
G_{\tau \tau'}(\omega) = V_{\tau \tau'} 
 + \sum_{k, k'} V_{\tau \tau'} \frac{\mid\bm{k},\bm{k'}\rangle \, Q_{\tau \tau'}\, \langle\bm{k},\bm{k'}\mid}{\omega-\epsilon_{\tau}(k)-\epsilon_{\tau'}(k') + i\varepsilon}
G_{\tau \tau'}(\omega) \;,
\label{bg}
\end{equation}
where $\tau, \tau'=n, p $ are isospin indices, $V_{\tau \tau'}$ denotes the bare NN interaction in a given NN channel, 
${\mid\bm{k},\bm{k'}\rangle \, Q_{\tau \tau'}\, \langle\bm{k},\bm{k'}\mid}$ is the Pauli operator which projects the intermediate 
nucleons states out of the Fermi sphere. In this way the Pauli exclusion principle is automatically satisfied.    
$\omega$ is the so-called starting energy which is given by the sum of energies of the interacting nucleons in a non-relativistic approximation.   
The single-particle energy $\epsilon_\tau(k)$ of a nucleon with momentum $k$ and mass $m_\tau$ is given by: 
\begin{equation}
       \epsilon_\tau(k) = \frac{\hbar^2k^2}{2m_\tau} + U_\tau(k) \ ,
\label{spe}
\end{equation}
where the single-particle potential $U_\tau(k)$ is the mean field felt by one nucleon due to the interactions with the other nucleons of 
the medium. 
In the BHF approximation, $U_\tau(k)$ is given by the real part of the $G_{\tau \tau'}$-matrix calculated  
on-energy-shell: 
\begin{equation}
U_\tau(k) =\sum_{\tau'=n,p}\ \sum_{k'\le k_{F_{\tau'}}} \mbox{Re} \ \langle \bm{k} \bm{k'}
\mid G_{\tau \tau'}(\omega=\omega^*) \mid \bm{k} \bm{k'}\rangle_A 
 \;,
\label{spp}
\end{equation}
where $\omega^*=\epsilon_\tau(k)+\epsilon_{\tau'}(k')$ and the sum runs over all neutron and proton occupied states and the matrix elements are antisymmetrized. 
In the solution of the Bethe--Goldstone equation, we have employed the so-called continuous choice \citep{jeuk+67,gra87} for the single-particle potential $U_\tau(k)$. 
It has been shown in Refs.\  \cite{song98,baldo00} that  
the contribution to the energy per particle $E/A$ from the diagrams coming from the three-hole-lines, is strongly minimized using this prescription. 
Consequently, a faster convergence of the hole-line expansion for $E/A$ is achieved\ \citep{song98,baldo00,baldo90} 
when compared to the so-called gap choice for $U_\tau(k)$ where the single particle potential are set to zero above the Fermi momentum. 

Eqs.\ (\ref{bg})--(\ref{spp}) are solved in a self-consistent way and then the energy per particle 
of the is calculated as:  
\begin{equation}
\frac{E}{A}(\rho,\beta)=\frac{1}{A}\sum_{\tau=n,p}\sum_{k \le k_{F_\tau} }
 \left(\frac{\hbar^2k^2}{2m_\tau}+\frac{1}{2} U_\tau(k) \right) \ .
\label{bea}
\end{equation}
From the energy per particle, all the other relevant quantities can be calculated using standard thermodynamical relations. 

\subsection{Inclusion of three-nucleon forces in the BHF approach}
Non-relativistic quantum many-body approaches are not able to reproduce the empirical saturation point of symmetric nuclear matter:  
$\rho_{0} = 0.16 \pm  0.01~{\rm fm}^{-3}$, $E/A|_{\rho_0} = -16.0 \pm 1.0~{\rm MeV}$. 
Several studies employing a large variety of different NN potentials have indeed shown that the saturation points lie inside a narrow band  
known in literature as Coester band \citep{coester70,day81}. The various models showed either a too large saturation density or a too small 
value for the energy per particle with respect to the empirical value. 
A similar behaviour has been also found for the binding energies of finite nuclei where the 
ground states turned out to be too large or too small when compared to the experimental ones. 
The inclusion of TNFs allows to improve the description of both SNM nuclear matter \citep{FP81,bbb97,apr98} and finite nuclei.   
In addition TNFs are very important in the case of $\beta$-stable nuclear matter to get an equation of state stiff enough to produce 
neutron star masses able to fulfill the limits put by the measured masses $M = 1.97 \pm 0.04 \, M_\odot$\ \citep{demo10} 
and $M = 2.01 \pm 0.04 \, M_\odot$\ \citep{anto13} of the neutron stars in PSR~J1614-2230 and PSR~J0348+0432 respectively.  

 
However in the BHF approach, as well as in almost all microscopic many body approaches, TNFs cannot be employed directly 
without approximation.  
This is because it would be necessary to solve very complicated three-body Bethe-Faddeev equations in the nuclear medium 
(Bethe--Faddeev equations) \citep{bethe65,rajaraman-bethe67}. Although this may be attempted in next future, for now this 
is a task beyond our possibilities. 
In order to bypass this problem, an average density dependent two-body force is built starting from the 
original three-body one. The average is made over the coordinates (including also spin and isospin degrees of freedom) of one of the 
three nucleons \citep{loiseau,grange89}.     

In the present work, we have used the in medium effective NN force derived by\ \cite{holt}  
which has the following structure:
\begin{eqnarray}
&&V_{eff}(\bm p,\bm q) = V_C + \bm \tau_1 \cdot \bm \tau_2\, W_C  \nonumber \\
&+& \left [V_S + \bm \tau_1 \cdot \bm \tau_2 \, W_S \right ] \bm \sigma_1 \cdot 
\bm \sigma_2  \nonumber \\
&+& \left [ V_T + \bm \tau_1 \cdot \bm \tau_2 \, W_T \right ] 
\bm \sigma_1 \cdot \bm q \, \bm \sigma_2 \cdot \bm q \nonumber \\
&+& \left [ V_{SO} + \bm \tau_1 \cdot \bm \tau_2 \, W_{SO} \right ] \,
i (\bm \sigma_1 + \bm \sigma_2 ) \cdot (\bm q \times \bm p) \nonumber \\
&+& \left [ V_{Q} + \bm \tau_1 \cdot \bm \tau_2 \, W_{Q} \right ] \,
\bm \sigma_1 \cdot (\bm q \times \bm p)\, \bm \sigma_2 \cdot (\bm q \times \bm p) \, . 
\end{eqnarray}
The subscripts on the functions  $V_i$, $W_i$ stand for central (C), spin (S), tensor (T), spin-orbit (SO) and quadratic spin-orbit (Q). 
(see\ \cite{holt} for the explicit expressions of these functions). 
This effective interaction can be obtained by averaging the original three-nucleon interaction 
$V_{3N}$ over the generalized coordinates of the third nucleon: 
\begin{equation}\label{eq:normord_singpart}
V_{eff} =
\text{Tr}_{(\sigma_3,\tau_3)} \int \frac{d \bm{p}_3}{(2 \pi)^3}
\, n_{\bm{p}_3} \, V_{3N} \, (1-P_{13}-P_{23})\, ,
\end{equation}
where 
\begin{equation}
  P_{ij} = \frac{1+\bm\sigma_i\cdot\bm\sigma_j}{2} \, \frac{1+\bm\tau_i\cdot\bm\tau_j}{2} \,
         P_{\bm p_i \leftrightarrow\bm p_j}
\end{equation}
are operators which exchange the spin, isospin and momentum variables of the nucleons $i$ and $j$. $n_{\bm{p}_3}$ is the Fermi 
distribution function at zero temperature of the "third" nucleon with momentum $\bm{p}_3$. Here we assume for $n_{\bm{p}_3}$ a step 
function approximation. 

\section{Results for nuclear matter}
\label{sec:3}

\begin{figure}[t]
\begin{center}
\includegraphics[width=0.5\textwidth]{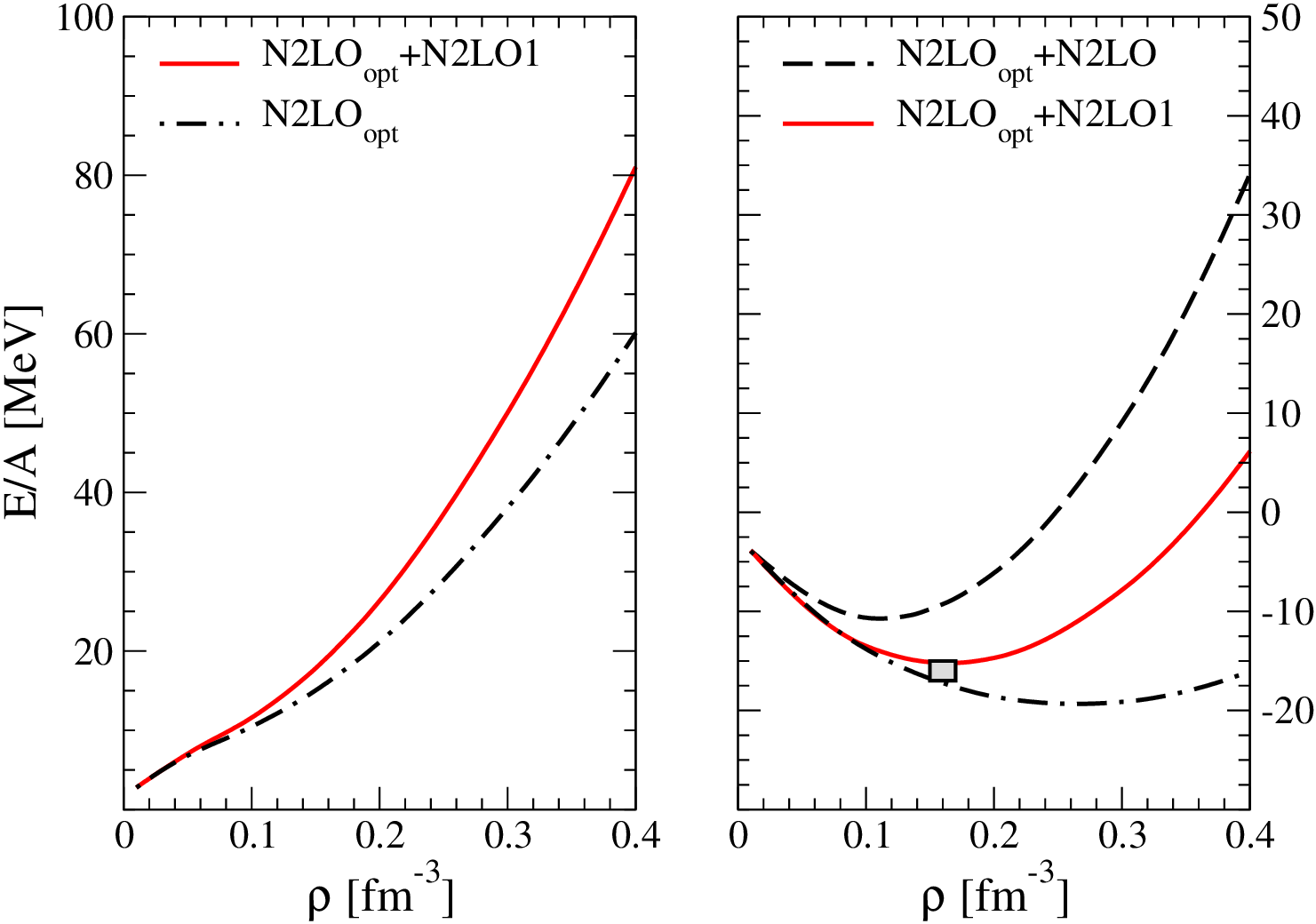}
\caption{(Color online) In the figure we show the energy per particle of pure neutron matter (left panel) and symmetric nuclear matter 
(right panel) as function of the nuclear density ($\rho$) for the two models described in the text. 
The empirical saturation point of nuclear matter  
$\rho_{0} = 0.16 \pm  0.01~{\rm fm}^{-3}$, $E/A|_{\rho_0} = -16.0 \pm 1.0~{\rm MeV}$ 
is represented by the grey box in the right panel. See text for details. } 
\label{fig1}
\vspace{0.5cm}
\end{center}
\end{figure}

In this section we discuss the results concerning the calculation of the energy per particle $E/A$ as 
a function of the nuclear density $\rho$, for pure neutron matter (PNM) and SNM using the two interaction models 
and the BHF approach described previously.  
In order to perform a partial wave expansion of the Bethe--Goldstone equation (\ref{bg}), we have made the usual angular average 
on the Pauli operator as well as on the energy denominator in the propagator \ \citep{gra87}.  
For each calculation, we have included all partial wave contributions up to a total 
two-body angular momentum $J_{max} = 8$. The contributions coming from higher partial waves are completely negligible.  
%
\begin{table*}
\caption{\label{tab2} Nuclear matter properties at saturation density ($\rho_0$) for the two models discussed in the text. 
In the first column of the table is reported the model name; in the other columns we give the saturation point of SNM,   
($\rho_0$), the corresponding value of the energy per particle ($E/A$), the symmetry energy ($E_{sym}$), 
the slope $L$ of $E_{sym}$ and the incompressibility $K_\infty$. 
All these values are referred to the saturation density ($\rho_0$) calculated for each model. } 
\vspace{0.2cm}
\begin{center} 
\begin{tabular}{cccccc}
\hline
Model & $\rho_0$(fm$^{-3}$) & $E/A$ (MeV) & $E_{sym}$ (MeV)  & $L$ (MeV) & $K_\infty$ (MeV) \\
\hline
N2LO$_{opt}$+N2LO1     & 0.163  & -15.20   & 34.38    &  79.01    & 222    \\
N2LO$_{opt}$+N2LO      & 0.110  & -10.72   & 24.03    &  35.70    & 134    \\
\hline
\end{tabular}
\end{center} 
\end{table*}
%
In Fig.\ \ref{fig1} we show the density behaviour of the energy per particle of PNM (left panel) and SNM (right panel) for both the models 
 considered in the present work. The dashed dotted lines in Fig.\ \ref{fig1} have been obtained using just 
the N2LO$_{opt}$ NN interaction without TNFs. 
We note that in the case of PNM employing either the N2LO or the N2LO1 TNF, the curve of the energy per particle does not change (red continuous line in left panel of Fig.\ \ref{fig1}). This happens because when performing the average 
of the TNF in pure neutron matter to get the effective density dependent two-body force $V_{eff}$ 
(see Eq.\ (\ref{eq:normord_singpart})), the terms containing the low energy constants $c_D$ and $c_E$ vanish for symmetry reasons 
(see\ \cite{logoteta16_D} for more details) while the other low energy constants 
$c_1$, $c_3$ and $c_4$, which take contribution to the average have the same values in the two models. 
Thus in PNM $V_{eff}$ is the same both for the N2LO1 and N2LO TNF. 
The effect of the TNF in both models is to produce a stiffer EOS. 
This is actually needed to improve the saturation point of SNM obtained using the sole NN interaction 
(black dashed dotted line in right panel of Fig.\ \ref{fig1}). 
In the latter case the saturation point turns out to be: 
$\rho_0 = 0.26$ ${\rm fm}^{-3}$ and $E/A|_0 = -19.23$ ${\rm MeV}$. 
Using the model N2LO$_{opt}$+N2LO1 a better nuclear matter saturation point is obtained: 
$\rho_0 = 0.163$ ${\rm fm}^{-3}$ and $E/A|_0 = -15.20$ ${\rm MeV}$. 
The empirical saturation point of SNM is represented by a grey box in Fig.\ \ref{fig1}. 
For the model N2LO$_{opt}$+N2LO the repulsion provided by the TNF, needed to reproduce the binding energies of light nuclei, is too strong in nuclear matter and the resulting curve of the energy per particle (black dashed line in right panel of Fig.\ \ref{fig1}) saturates at a too small density comparing to the empirical one. 
For the model N2LO$_{opt}$+N2LO the saturation point of SNM is 
$\rho_0 = 0.110$ ${\rm fm}^{-3}$ and $E/A|_0 = -10.72$ ${\rm MeV}$. 
The values of the saturation density and energy per particle at saturation for the two models considered are 
reported in Tab.\ \ref{tab2}.     

The energy per particle of asymmetric nuclear matter, which is essential to describe neutron stars, can be calculated with very good 
accuracy using the so called parabolic approximation \citep{bl91}:  
\begin{equation}
      \frac{E}{A}(\rho, \beta) = \frac{E}{A}(\rho, 0) + E_{sym} (\rho) \beta^2 \,,
\label{parab}
\end{equation}
where $E_{sym}(\rho)$ is the nuclear symmetry energy \citep{epja50} and $\beta$ is the asymmetry parameter defined in the previous section. 
Using Eq.\, (\ref{parab}), the symmetry energy can be obtained from the difference between the energy 
per particle of PNM ($\beta = 1$) and SNM ($\beta = 0$).

\begin{figure}[t]
\begin{center}
\includegraphics[width=0.5\textwidth]{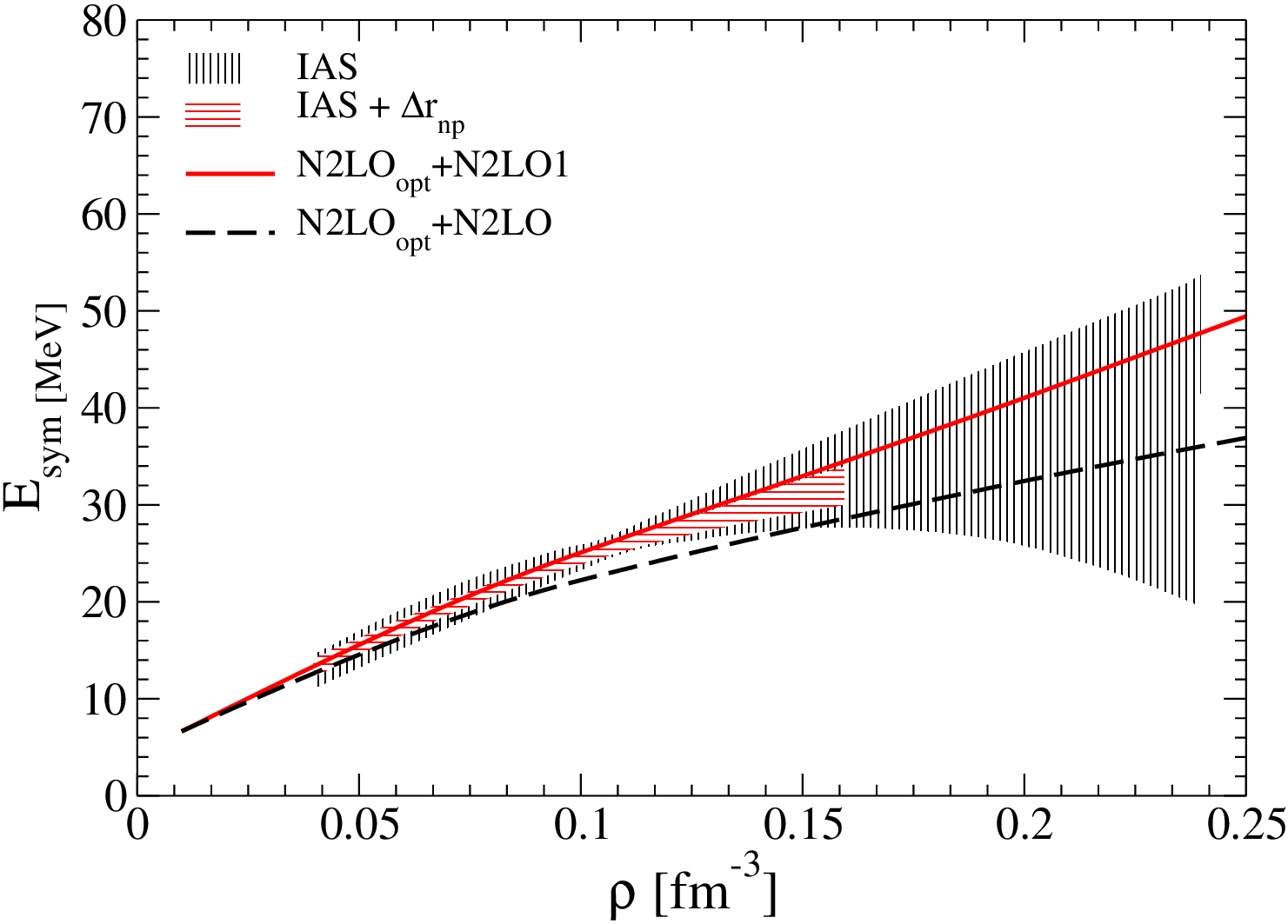}
\caption{(Color on line)  The nuclear symmetry energy is shown as a function of the nucleonic density for the two  
interaction models used in the present work. 
The constraints on the symmetry energy obtained by\ (\cite{dan14})  
using the excitation energies of isobaric analog states (IAS) in nuclei are represented by the black-dashed band, labeled IAS.   
The smaller region covered by the red-dashed band labeled IAS+$\Delta r_{np}$ (\cite{roca13}) are additional constraints provided by 
 the data analysis of neutron skin thickness ($\Delta r_{np}$) of heavy nuclei. 
}
\label{fig_esym}
\end{center}
\end{figure}

In Tab.\ \ref{tab2} we show the values of the symmetry energy and the so called slope parameter $L$ defined as: 
\begin{equation}
       L = 3 \rho_{0} \frac{\partial E_{sym}(\rho)}{\partial \rho}\Big|_{\rho_{0}} 
\label{slope}
\end{equation} 
at the calculated saturation density $\rho_0$ (second column in Tab.\ \ref{tab2}) 
for the two interaction models considered in the present paper.  
We note that the values of $E_{sym}(\rho_0)$ and $L$ calculated with model N2LO$_{opt}$+N2LO1 are in a good agreement 
with those obtained by other calculations based on the BHF approach including two- and three-body forces  
(see e.g.\ \citep{ZHLi06,li-schu_08}) and with the values derived from different experimental data as discussed by\ \cite{latt14}. 
Our second model instead underestimates both the values of $E_{sym}$ and $L$.         

The incompressibility $K_\infty$ of SNM calculated at saturation density is given by: 
\begin{equation}
       K_\infty = 9 \rho_{0}^2 \frac{\partial^2 E/A}{\partial \rho^2}\Big|_{\rho_{0}} \,.
\label{incom}
\end{equation}
The value of the incompressibility $K_\infty$ can be obtained analyzing experimental data of giant monopole resonance (GMR) 
energies in medium and heavy nuclei. Such analysis performed first by \cite{blaizot76}, provided the value $K_\infty = 210 \pm 30$ 
${\rm MeV}$. The refined analysis of\ \cite{shlo06} gave instead the value: $K_\infty = 240 \pm 20$~${\rm MeV}$.   
Recently\ \cite{stone10} on the basis of a re-analysis of GMR data found: $250$ ${\rm MeV}$$<K_\infty<$ $315$ ${\rm MeV}$.   
In the last column of Tab.\ \ref{tab2} we have reported the incompressibility $K_\infty$, at the calculated saturation point $\rho_0$ 
for the two models considered in the present work.    
Model N2LO$_{opt}+N2LO1$ is in very good agreement with the value of $K_\infty$ predicted 
by\ \cite{blaizot76} and\ \cite{shlo06}. It should be noted that the value of $K_\infty$ is a very important quantity not only for nuclear physics but also for astrophysics. It has been shown indeed that $K_\infty$ is strongly correlated to the physics of supernova explosions 
and neutron star mergers.

\begin{figure}[t]
\centering
\includegraphics[width=0.5\textwidth]{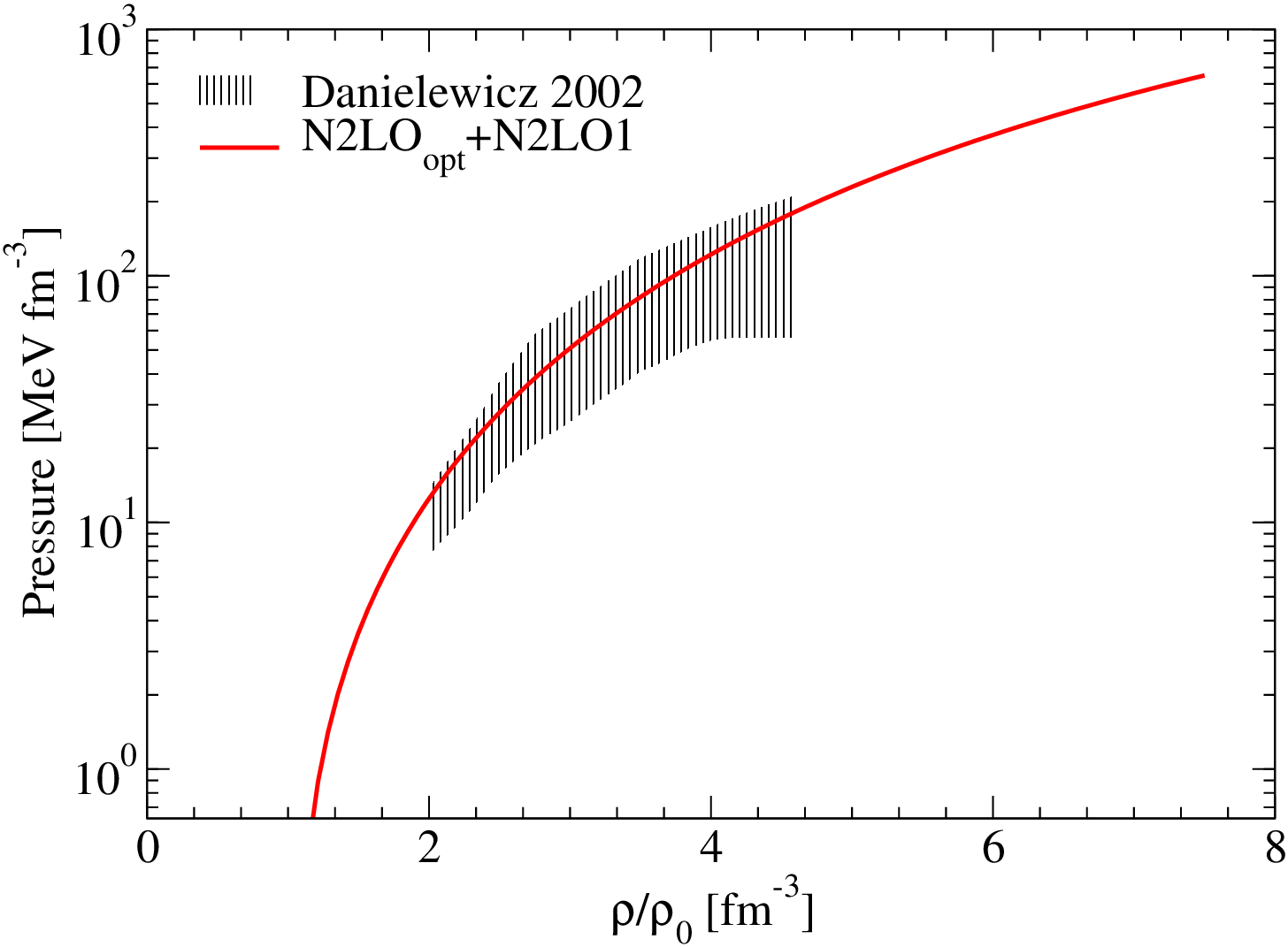}
\caption{(Color online) Pressure of SNM as a function of the nucleonic density $\rho$ 
(in units of the empirical saturation density $\rho_0=0.16$ ${\rm fm}^{-3}$) for the model N2LO$_{opt}$+N2LO1. 
The black hatched area represents the region for SNM which is consistent
with the constraints provided by collision experiments between heavy nuclei (\cite{Dan2002})}
\label{fig31}
\end{figure}


Another important constraint that should be fulfilled by a good nuclear matter EOS, concerns the behaviour of the pressure 
of SNM as function of the nucleonic density. Such constraints are provided by experiments of 
collisions between heavy nuclei. In such experiments matter is compressed up to $\sim 4 \rho_0$ and it is 
therefore possible to extract important informations about the behaviour of the EOS at densities larger than normal 
saturation density ($\rho_0=0.16$ ${\rm fm}^{-3}$)). 

The black hatched area in Fig.\ \ref{fig31} is the region in the pressure--density plane for SNM determined by \cite{Dan2002}, 
performing several numerical simulations able to reproduce the measured elliptic flow of matter in the collision experiments 
between heavy nuclei.  

In the same figure, we show the pressure of SNM for the N2LO$_{opt}$+N2LO1 (red continuous line) model
 obtained from the calculated energy per nucleon and using the standard thermodynamical relation:  
\begin{equation}
         P(\rho)  = \rho^2 \frac{\partial (E/A)}{\partial \rho}\Big|_{A} \ . 
\label{PresSNM}
\end{equation}  
Our results are fully consistent with the empirical constraints given by\ \cite{Dan2002}.

\section{Neutron star structure} 

We next apply the model N2LO$_{opt}$+N2LO1, 
which reproduces various empirical nuclear matter properties at the saturation density (Tab.\ \ref{tab2}),    
to calculate the structure of neutron stars.     

The composition of the inner core of neutron stars cannot be completely determined by data from observations and 
therefore different scenarios are currently under consideration. 
The appearance of hyperons\ \citep{hyp1,Isaac2011} or the transition to a phase with deconfined quarks 
(quark matter) \citep{gle96,bom09,logoteta12,bomb13,logoteta13} are among the most admissible possibilities.

In this work we want mainly to concentrate on the simplest case of pure nucleonic matter with the aim 
to establish if the modern chiral nuclear interactions considered here, can provide an EOS which is able 
to fulfill the constraints put by observational data on neutron stars properties.     
This first check represents a mandatory step before to explore more sophisticated possibilities with 
additional feasible degrees of freedom.  
We point out however that allowing for a quark deconfinement phase transition and considering the possible       
existence of a second branch of compact stars (quark stars) with "large" masses compatible with present mass measurements, i.e. within the so-called two families scenario \citep{be03,bo04,bombaci16,drago16}, 
is not necessary that the neutron star branch reproduces the limit of two solar masses.    

We also report a calculation of the EOS that includes, in addition to nucleons, hyperonic degrees of freedom 
and in particular the presence of $\Lambda$ and $\Sigma^-$ hyperons. These are in fact the first hyperon 
species expected to appear in microscopic calculations of neutron star matter \citep{hyp1,Isaac2011, schulze06}.      
We thus consider also the so-called hyperonic stars.   

In order to determine the mass-radius (M(R)) and mass-central density (M($\rho_c$)) relations for non rotating 
neutron stars one needs first to calculate the $\beta$-stable EOS of the system.  
The composition of $\beta$-stable stellar matter is determined by the relations between the chemical potentials 
of the various constituent species. 
In this paper we consider neutrino free matter 
($\mu_{\nu_e} = \mu_{\bar{\nu}_e} = \mu_{\nu_\mu} = \mu_{\bar{\nu}_\mu}$) 
in the general case of matter if matter with hyperons. We have:    
\begin{equation}
\mu_n-\mu_p=\mu_{e^-} \;, \ \ \ \ \ \ \ \mu_{e^-}=\mu_{\mu^-}, \\   
\label{beta1}
\end{equation}
\begin{equation}
\mu_\Lambda=\mu_n \;, \ \ \ \ \ \ \ \mu_{\Sigma^-}=\mu_n+\mu_{e^-}. \\
\label{beta1y}
\end{equation}
In Eqs.\ (\ref{beta1}) and (\ref{beta1y}) $\mu_n$, $\mu_p$, $\mu_\Lambda$, $\mu_{\Sigma^-}$, $\mu_{e^-}$ and $\mu_{\mu^-}$ are chemical potentials of neutron, proton, $\Lambda$, $\Sigma^-$, electron and muon. 
Finally charge neutrality requires:  
\begin{equation}
\rho_p=\rho_{\Sigma^-}+\rho_{e^-}+\rho_{\mu^-}
\label{beta2}
\end{equation}
The various chemical potentials of baryons ($B=n,p,\Lambda,\Sigma^-$) and leptons ($l=e^-,\mu^-$) are determined through: 
\begin{equation}   
\mu_B=\frac{\partial \epsilon}{\partial \rho_B}\;, \ \ \ \ \mu_l=\frac{\partial \epsilon}{\partial \rho_l} \;
\label{beta3}
\end{equation}
where $\epsilon = \epsilon_{N}+\epsilon_{Y}+\epsilon_L$ is the total energy density which sums up the 
the nucleonic contribution $\epsilon_{N}$, the hyperonic one $\epsilon_{Y}$ and the leptonic one $\epsilon_L$. 
The nucleonic contribution $\epsilon_{N}$ has been calculated using the N2LO$_{opt}$+N2LO1 nuclear interaction     
and the thermodynamical relation $\epsilon_{N}=\rho\, E/A(\rho,\beta)$,  
with the energy per particle $E/A(\rho,\beta)$ of asymmetric nuclear matter calculated in BHF approximation 
and employing the parabolic approximation \citep{bl91}.
For the hyperonic contribution $\epsilon_{Y}$ we have used the parametric form of the BHF energy per particle 
of asymmetric hyperonic matter provided by\ \cite{shulze16} and obtained using the nucleon-hyperon (NY) and 
hyperon-hyperon (YY) interactions. More specifically\ \cite{shulze16} used the NY Nijmegen soft core NSC08b 
potential \citep{NSC08b} supplemented with the new YY Nijmegen soft core NSC08c potential\ \citep{NSC08c}.   
We note that these interactions have been derived following the scheme of traditional meson exchange theory 
and not in the framework of ChPT. However they provide an accurate description of the available hypernuclear 
data\ \citep{NSC08b}.         

We have then self-consistently solved the equations (\ref{beta1}), (\ref{beta1y}), (\ref{beta2}), (\ref{beta3}) 
as function of the total baryonic density $\rho=\rho_n+\rho_p+\rho_\Lambda+\rho_{\Sigma^-}$ and obtained 
the EOS for $\beta$-stable hyperonic matter with nucleons, hyperons, electrons and muons ($\mu^{-}$).   

\begin{figure}[t]
\centering
\includegraphics[width=0.5\textwidth]{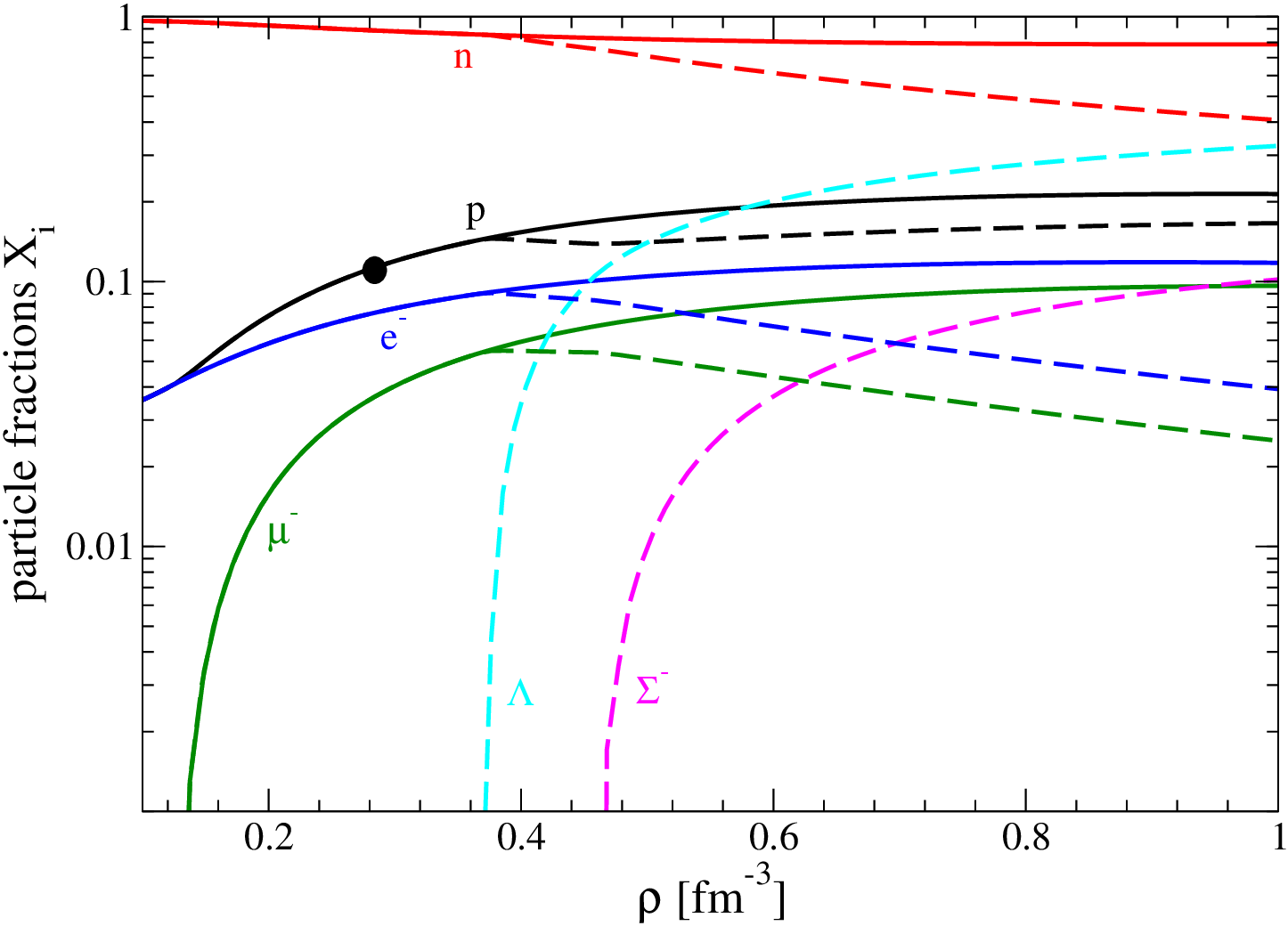} 
\caption{(Color online) Particle fractions in $\beta$-stable neutron star matter for model N2LO$_{opt}$+N2LO1.
The continuous lines (dashed lines) refer to particle fractions in the case of $\beta$-stable 
nucleonic matter (hyperonic matter). }
\label{comp}
\end{figure}

The composition of $\beta$-stable nucleonic matter is shown by the continuous lines in Fig.\ \ref{comp}.  
The black circle on the black line which represents the proton fraction, marks the density threshold  
for the direct URCA processes $n \rightarrow p + e^- + \bar{\nu}_e\,$, 
$~ p + e^-  \rightarrow n +  \nu_e\,,$ (\cite{dURCA}).   
In our model this threshold is  $\rho_{DU} = 0.339\ \rm{fm}^{-3}$  which corresponds to a neutron star mass 
$M(\rho_{DU}) = 0.97\ M_\odot$.   
The dashed lines in Fig.\ \ref{comp} represent the results of the solution of the $\beta$-equilibrium equations 
for hyperonic matter with $\Lambda$ and $\Sigma^-$ hyperons.  
The $\Lambda$ hyperon is the first hyperonic species to appear at a density around $0.37$ {\rm fm}$^{-3}$ 
while the $\Sigma^-$ hyperon appears at density of $0.47$ {\rm fm}$^{-3}$. 
This behaviour is a new feature of modern NY interactions which find a much more repulsive contribution    
in the N$\Sigma^-$ channel to the total energy density.  The same trend has been also found by recent NY 
interactions derived in ChPT by\ \cite{haidenb}. Such a repulsion leads to the appearance of 
the  $\Lambda$ hyperon before the $\Sigma^-$ one contrarily to the predictions of older NY  interaction 
models\ \citep{schulze06}.    

\begin{figure*}[t]
\centering
\includegraphics[width=0.76\textwidth]{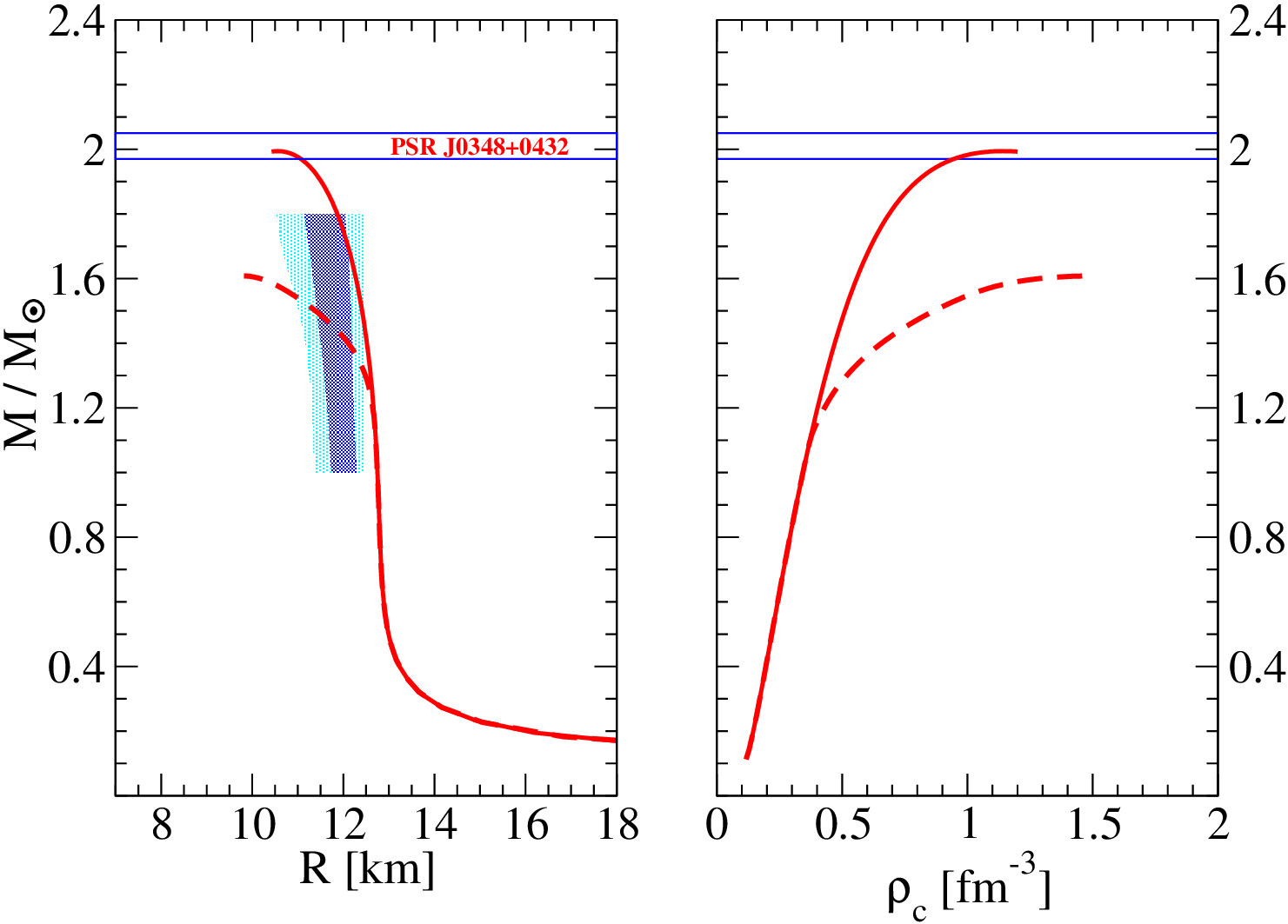}
\caption{(color online) Mass-radius ($M(R)$) (left panel) and mass-central density ($M(\rho_c)$) (right panel) relationships for the models described in the text. 
The continuous lines refer to the calculation performed considering the EOS containing only nucleonic degrees of freedom 
while the dashed lines have been obtained including also the $\Lambda$ and the $\Sigma^-$ hyperons in the calculation.   
The hatched region in the left panel represents the the mass-radius constraints obtained by\ 
\cite{steiner10,steiner13}). The strip with boundaries marked with blue lines stands 
for the measured mass $2.01 \pm 0.04 M_\odot$ \citep{anto13} of the neutron stars in PSR J0348+0432.}
\label{fig5}
\end{figure*}

In order to calculate the neutron stars structure, we have numerically solved the equations for hydrostatic 
equilibrium in general relativity \citep{Tol39,OV39}.  
For nucleonic density smaller than $0.08$ fm$^{-3}$ we have matched our EOS models of the core with the \cite{NV73} and 
 Baym--Pethick--Sutherland \citep{bps}) EOSs which model neutron stars crust.  
 
In Fig.\ \ref{fig5} we show the results of our calculations. In the left (right) panel we plot the mass-radius (mass-central density)  relations for our models.  
Referring now to the left panel in Fig.\ \ref{fig5}, the hatched regions are constraints derived from     
the analysis of observational data of both transiently accreting and bursting X-ray sources   
obtained by \cite{steiner10,steiner13}.    
We note the maximum mass $M_{max} = 1.99$ $M_{\odot}$ obtained for nucleonic stars, i.e. for the EOS model 
including only nucleons (continuous line in Fig.\ \ref{fig5}), is compatible with present neutron star mass measurements and in particular with the measured mass $2.01 \pm 0.04 M_\odot$ \citep{anto13} of the neutron star  
in PSR J0348+0432 (strip with boundaries marked with blue lines in Fig.\ \ref{fig5}). 
In addition our results are also in rather good agreement with the empirical constraints on the mass-radius  relationship reported in \cite{steiner10,steiner13}. 
We note however that presently there is no general agreement on neutron star radii measurements due to 
the large uncertainties in the techniques used to extract this quantity.  
For instance small stellar radii in the range of $9-12\ \rm{km}$ \citep{gui13} are found considering informations from spectral analysis 
of X-ray emission from quiescent X-ray transients in low-mass binaries (QLMXBs). Larger radii around $16\ \rm{km}$ are instead obtained 
considering data on neutron stars with recurring powerful bursts.  
However these last measurements are subject to large uncertainties \citep{pou14}.   
In a recent work \cite{lattimer16} suggests that neutron star radii should lie in the range between 
$10.7-13.1\ \rm{km}$.    

The red dashed lines in Fig.\ \ref{fig5} represent the mass-radius (left panel) and mass-central density (right panel) relations for hyperonic stars (i.e. for the EOS model including hyperons in addition to nucleons). 
In this case there is a sizable decrease of the stellar maximum mass down to $M_{max} = 1.6$ $M_\odot$,  
a value which is incompatible with measured neutron star masses. 
This outcome is caused by the softening of the EOS due to the presence of hyperons in the stellar core 
\citep{schulze06,Isaac2011,log12}.    

This difficulty to reconcile the measured masses of neutron stars with the seemingly unavoidable presence 
of hyperons in their interiors is called {\it hyperon puzzle} \citep{lonard,bombHYP2015,chatt2016} in  
neutron stars.   
This unsolved puzzle is currently the subject of several investigations and various possible solutions have been  proposed. Some researches pointed out the importance of taking into account the effect of hyperonic three-body 
forces between nucleons and hyperons \citep{lonard,Isaac2011,chatt2016}, 
while other investigations \citep{bombaci16,drago16} underline the possibility for a phase transition to  
quark matter at large baryonic density and the existence of a second branch of compact stars (quark stars) with "large" masses compatible with present mass measurements.  
Finally we emphasize that also the two-body YY interaction can play a role in solving the hyperon puzzle.  
In fact, as shown by\ \cite{schulze06}, the new NSC08c YY interaction makes the EOS stiffer and allows to increase 
the maximum mass of about $0.25$ $M_\odot$ with respect to the case when only NN and NY interactions are taken into account to describe the two-body baryon-baryon interactions. 

\begin{table}
\caption{\label{tab4} Mass (in unit of solar mass $M_\odot=1.989\times10^{33} \rm{g}$), corresponding radius (in $\rm{km}$) and central density (in $\rm{fm^{-3}}$) for the neutron star configuration corresponding to the maximum masses of Fig.\ \ref{fig5}.}
\small
\vspace{0.2cm}
\begin{tabular}{cccc}
\hline
\hline
Model                          & $M$ ($M_\odot$) & $R$ (km) & $\rho_c$ (fm$^{-3}$)  \\
\hline
N2LO$_{opt}$+N2LO1             & 1.99            & 10.52    & 1.13                  \\ 
N2LO$_{opt}$+N2LO1+NY+YY       & 1.60            & 9.86     & 1.50                  \\ 
\hline
\hline
\end{tabular}
\end{table}

The properties of the maximum mass configuration for our models of nucleonic and hyperonic stars are reported 
in Tab.\ \ref{tab4}. These results are in good agreement with other calculations based on microscopic approaches. 
Concerning this point it is interesting to note that our present findings are very similar to those reported in\ \cite{taranto13} where nuclear matter properties and $\beta$-stable EOS have been obtained using the BHF approach and employing two- and three-body forces based on the meson-exchange theory. In addition our results are in good accord with those in\ \cite{bombaci18} where the neutron stars  structure was described adopting chiral potentials calculated in the so called $\Delta$-full theory both at two- and three-body level. 
 Such agreement provides an independent way to check the correct behaviour of the interactions used in the present work at large baryonic density. We note indeed that the interactions derived in ChEFT are characterized by a low-momentum expansion and therefore can be trusted up to baryonic densities for which the Fermi momentum is of the order of magnitude of the cutoff set in the regulator function. At larger densities the EOS should be extrapolated or an accurate analysis of convergence of the many-body calculation has to be properly accounted for. We note that for neutron stars these considerations are mandatory because the maximum density reached in the core can be even larger than $1$ fm$^{-3}$ (see Tab.\ \ref{tab4}).  


\begin{figure}[t]
\centering
\includegraphics[width=0.45\textwidth]{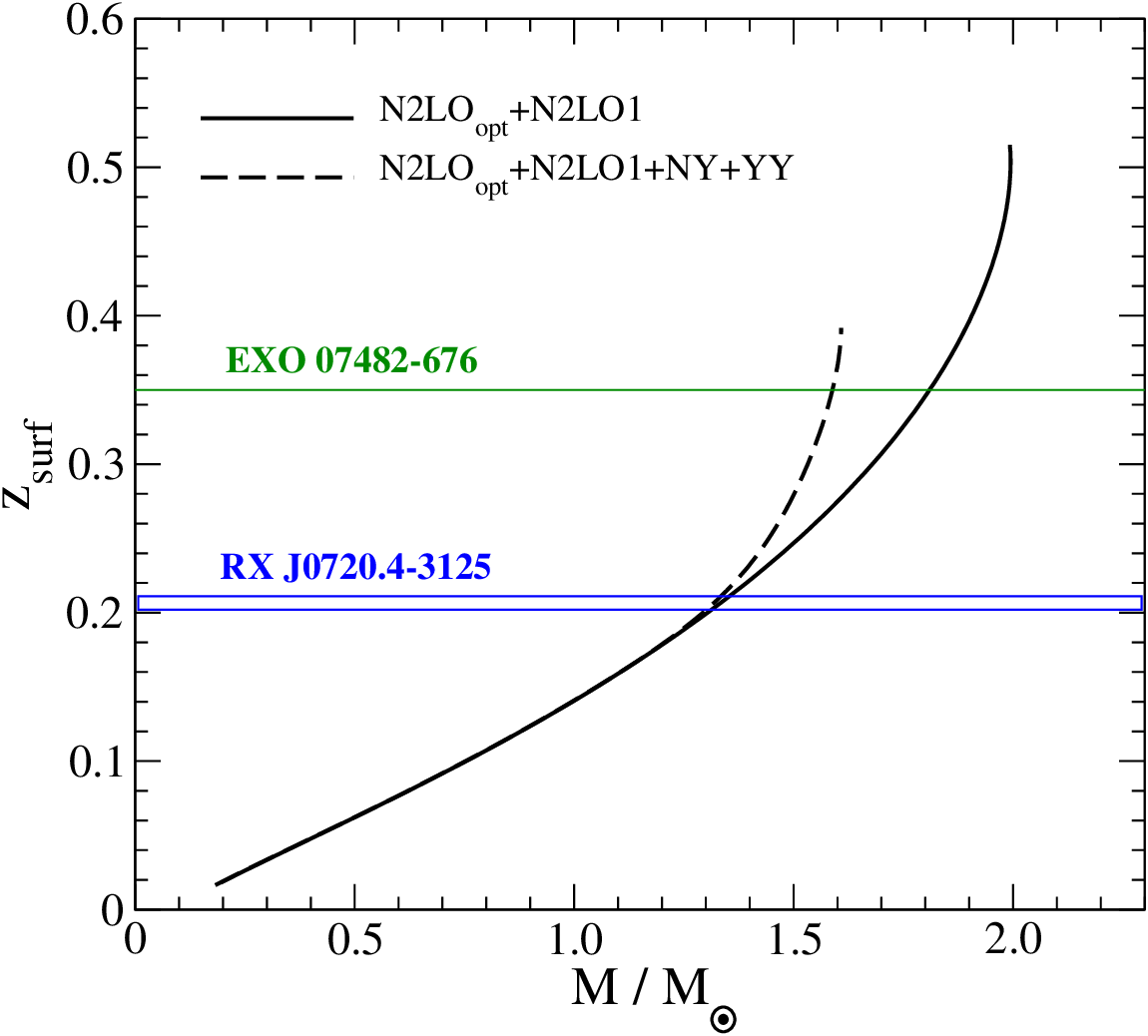}
\caption{(color online) Gravitational redshift calculated at the neutron star surface as a function of the stellar gravitational mass for the two EOS models used in our work. 
The horizontal lines stand for the measured gravitational redshift $z=0.35$ for the X-ray bursts source in the 
low-mass X-ray binary EXO\,07482$-$676\ (\cite{cot}) and $z=0.205_{-0.003}^{+0.006}$ for the isolated neutron star RX\,J0720.4$-$3125\ 
\cite{hamb}.} 
\label{fig6}
\end{figure}

The gravitational redshift of a signal emitted from the stellar surface is given by:  
\begin{equation} 
   z_{surf} = \Bigg( 1 - \frac{2 G M}{c^2 R}\Bigg)^{-1/2} - 1 \,.  
\label{zsurf}
\end{equation} 
The measurements of $z_{surf}$ of spectral lines can provide a direct information on the neutron star  
compactness parameter: 
\begin{equation}
     x_{GR} = \frac{2 G M}{c^2 R} \;.
\label{xgr}
\end{equation}
and therefore on the EOS of neutron star matter. 
The calculation of the surface gravitational redshift for our two EOS models is shown in  
Fig.\ \ref{fig6}. The two horizontal lines in the same figure  
stand for the measured gravitational redshift  
$z = 0.35$ for the X-ray bursts source in the low-mass X-ray binary EXO\,07482$-$676 \citep{cot} and  
$z=0.205_{-0.003}^{+0.006}$ for the isolated neutron star RX\,J0720.4$-$3125 \citep{hamb}.


\section{Summary} 

We have investigated the behaviour and the properties of $\beta$-stable nuclear matter using two microscopic 
models based on nuclear hamiltonians obtained from ChPT at the N2LO, in the framework of many-body BHF approach. 
In particular we have used, the non local NN chiral potential derived by\ \cite{ekstrom1}  
which is able to reproduce the NN scattering data with a $\chi^2/datum\sim\ 1$. 
In order to get a good description of nuclear matter at saturation density we have included in our calculation 
also a TNF consistently calculated at the same order of ChPT.  
Concerning the TNF, we have explored two different parametrizations: 
the first one (N2LO) fitted to reproduce binding energies of light nuclei while the second one (N2LO1) 
fitted to reproduce a good saturation point of symmetric nuclear matter.     
We have shown that in the first case it was not possible to reproduce also good properties of nuclear matter at saturation density. 
For the second case we have shown that once the saturation point of SNM was well reproduced, 
other nuclear matter properties at the saturation density were also well determined. 
We have later calculated the EOS for $\beta$-stable nuclear matter for our best model, namely the 
N2LO$_{opt}$+N2LO1 one, and determined the neutron stars structure. 
We have found that the maximum mass obtained is compatible with the present measured neutron star masses.  
In addition we have found that the mass-radius relation for nucleonic stars is in a quite good  
agreement with the mass-radius constraints determined by\ \cite{steiner10,steiner13}. 
Finally we have extended our EOS model to include hyperons and we have thus calculated the corresponding 
hyperonic star properties. 
Confirming the results of previous studies, e.g. \citep{schulze06,Isaac2011,lonard,chatt2016}, 
we have found that the inclusion of hyperons leads to a substantial reduction of the value of the maximum mass 
which turns out to be not compatible with measured neutron star masses.   
This so-called hyperon puzzle is one of the hottest topics in neutron star physics  
which is stimulating copious experimental and theoretical research in hypernuclear physics. 

Several extensions of the present model to include hyperonic three-body forces and quark degrees of freedom are 
indeed under consideration. 
In addition the inclusion of thermal effects necessary for application to supernova explosions and consistent neutron star merger simulations  are also in development.     

\section*{Acknowledgments}
This work has been partially supported by ``NewCompstar'', COST Action MP1304.


\end{document}